# A New Assessment Statement for the Trinity Nuclear Test, 75 Years Later

Hugh D. Selby, Susan K. Hanson, Daniel Meininger, Warren J. Oldham, William S. Kinman, Jeffrey L. Miller, Sean D. Reilly, Allison M. Wende, Jennifer L. Berger, Jeremy Inglis, Anthony D. Pollington, Christopher R. Waidmann, Roger A. Meade, Kevin L. Buescher, James R. Gattiker, Scott A. Vander Wiel, and Peter W. Marcy

Los Alamos National Laboratory
Los Alamos, NM 87545

**Abstract**: New measurement and assessment techniques have been applied to the radiochemical re-evaluation of the Trinity Event. Thirteen trinitite samples were dissolved and analyzed using a combination of traditional decay counting methods and the mass spectrometry techniques. The resulting data were assessed using advanced simulation tools to afford a final yield determination of 24.8 ± 2 kilotons TNT equivalent, substantially higher than the previous DOE released value of 21 kilotons. This article is intended to complement the work of Susan Hanson and Warren Oldham, seen elsewhere in this issue.[1]

## I. Introduction

The Trinity nuclear test was fired on July 16, 1945. Its sudden and tremendous power was unlike anything humankind had ever witnessed. Virtually every aspect of the test was developed from whole cloth to study the unknown. The physics, engineering, metallurgy, chemistry, and fabrication that led to the Trinity device and the speed at which they were accomplished were unparalleled at the time and arguably remain so today.

This technical story has captivated scientists and non-scientists alike since that fateful July day. Over the years, no nuclear test has been as thoroughly studied. A simple example of the interest and fascination with Trinity is the fact that trinitite is readily purchased for conducting experiments and pedagogical demonstrations today. Such studies continue to fuel the steady stream of Trinity-based literature, 75 years after the test. Yet the yield of Trinity itself has been surprisingly uncertain and has changed several times in the decades between the test and this report.[2,3,4] These yields range from about 18 kt to 21 kt TNT equivalent (inside the uncertainty estimate of ~20% 1σ offered by the original Los Alamos Radiochemistry Group). The changes were driven largely by the desire to reexamine Trinity through the lens of more advanced measurement technology for example, or more detailed evaluation models. Regardless of motivation, all were limited by the developmental nature of the original dataset generated in 1945.

## II. Radiochemistry

How can this be? How can one of the most lavishly diagnosed nuclear tests, with some of the greatest minds characterizing its performance, not be known to better than ca. 20%? There are many answers to the question, but they can be reduced to two main themes: sampling and measurement technology. The challenges these two problems posed is best explained in the context of the Radiochemistry Group's origin.

The Radiochemistry Group was created for the express purpose of determining the efficiency of the Trinity core, or pit, through measurements of residual plutonium fuel and fission fragments. The ratio of these two classes of detonation products is proportional to the efficiency of the plutonium, which is derived from the following expressions.

$$Pu^0 = Pu^{res} + F \qquad (1)$$

Here, $Pu^0$ is the number of ingoing $^{239}$Pu atoms in the pit (about 6 kg or 1.15×10$^{25}$ atoms for Trinity, we ignore minor isotopes for reasons discussed below), $Pu^{res}$ is the unburned plutonium following the test, and $F$ is the number of fissions that occurred (alternatively, the number of original plutonium atoms that were converted to fission fragments through the fission process). Therefore efficiency can be estimated as the ratio of post-test fissions and ingoing plutonium:

$$E_{Pu} \sim \frac{F}{Pu^0} \qquad (2)$$

Substituting the definition of Pu$^0$ from Equation 1:

$$E_{Pu} = \frac{F}{Pu^{res}+F} \qquad (3)$$

Finally, combining efficiency with the known mass affords plutonium's explosive equivalent yield, as seen in Equation 4.

$$Y_{Pu} = M_{Pu} \times E_{pu} \times 18.2 \; kt/kg_{pu} \qquad (4)$$

Equation 4 conveys the concept of burning a known mass of fuel with a given efficiency, much like one might calculate the energy release of a mass of burning hydrocarbon. To determine the above conversion constant, 18.2 kt/kg, one first converts $M_{Pu}$ to the number of atoms of $^{239}$Pu (Trinity's ingoing isotopic profile contained less than 1% of $^{238,240,241,242}$Pu, hence energy release from fission in these isotopes was negligible):

$$N_{^{239}Pu} = grams\ ^{239}Pu \times \frac{1\ mol\ ^{239}Pu}{239.05\ grams\ ^{239}Pu} \times \frac{6.022 \times 10^{23}\ atoms}{mol} \quad (5)$$

The efficiency $E_{Pu}$ is dimensionless and is expressed in terms of fissions per atom of $^{239}$Pu. The total number of fissions is trivially converted to yield in equivalent kilotons of TNT detonation through multiplication by an appropriate energy release/fission value. We used the value of $7.216 \times 10^{-24}$ kt/f (or 188.4 MeV/f) $^{239}$Pu in our recently reported study, for example,[5] taken from Madland, which is in ENDF/B-VII.1 (see Ref. 6, Table XXVII, p. 2954, the prompt energy release at fast neutron incident energies for fission fragments, neutrons and gamma-rays).

Equation 3 indicates the type of measurements the Los Alamos Radiochemistry Group made.[7] Peak-yield fission product species such as $^{97}$Zr, with high specific activity and high relative production rates (near the peaks of the bimodal fission fragment distribution curve) were excellent proxies for F in the samples. These fragments were strong beta emitters and readily measured in purified samples. Plutonium (Pu$^{res}$) was quantified with alpha counting methods using aliquots subsampled specifically for the purpose.

A great effort was made to develop the exacting dissolution and subsequent separation chemistries required to make the alpha- and beta-emitter measurements. It is a testament to the Radiochemistry Group that modern radiochemical methods – within and external to the weapons science endeavor – are remarkably similar to those used in 1945. What has changed tremendously is the technology employed to quantify the various radioactivities discussed here. In 1945, the best practical method to quantify beta decay was the use of relatively unstable Geiger-Müller (GM) tubes. These required constant refurbishment and calibration to known quantities of a given beta-emitting species and their corresponding proportionality to fission events.[8] Indeed, one aspect of the non-nuclear "100 Ton Test" that preceded Trinity was to ascertain how well the group could quantify dispersed amounts of known beta activity using GM tubes.[9] Lower activity species (e.g., low-production neutron activation products or isotopes with long half-lives) were exceedingly difficult to quantify. Many such species were simply not measured in 1945.

Alpha counting was somewhat more robust and relied mainly on Frisch-grid detectors. The low interference rates from alpha emitters in a sample relative to plutonium were a major factor in the comparative ease of plutonium quantification. Still, the sample had to be extremely pure, with minimal residual fission activity remaining. This aspect of plutonium measurement was not really in hand until just after the 100 Ton Test. Taken together, these challenges (and many more) to producing measurements of Pu$^{res}$ and F illustrate the technological contribution to the uncertainty in the original yield estimate of Trinity.

The second factor that contributed to Trinity's uncertainty was the nature of the samples that would be collected. It was well understood at the time that, in order for Equation 3 to be true, samples had to contain amounts of Pu$^{res}$ and F that represented the actual plutonium efficiency. It was also known that the vapor-phase and condensation chemistries of the various fission fragments would vary significantly relative to plutonium. Consequently, much of the development in the Radiochemistry Group focused on determining which species condensed from the fireball with plutonium, and how to collect samples that contained the representative mixture. Two exemplars were the $^{95,97}$Zr isotopes, which were found to behave in a very similar chemical manner to plutonium. Knowing the chemical compatibility of plutonium and zirconium was only half the battle – it remained to be seen what kind of samples would contain the correct mixture. The sampling question was also part of the purpose of the 100 Ton Test, but no completely satisfactory method was produced before Trinity. Inconveniently, the nuclear detonation had not yet produced samples for method development! Many different techniques were ultimately used to collect samples, including the famous lead-lined tank, rockets, and fallout trays.[10] The various sampling methods unsurprisingly produced varying mixtures of Pu$^{res}$ and F. Careful consideration was given to selecting samples for yield determination without unduly biasing them. The end result was still ~20% variability in efficiency $E_{pu}$ from sample to sample, making sampling the largest contributor to Trinity's uncertainty estimate. As nuclear testing proceeded post-war, sampling remained a significant concern for radiochemical yield determination and required years of research to optimize.

Given the difficulties presented by early measurement technologies and sub-optimal samples, it is noteworthy how close the original yield estimate (18 kt) is to more modern assessments (21 kt). The accuracy also speaks volumes to the impressively complete understanding of condensation phenomenology the original Radio-chemistry Group developed for their first yield assessment. Nevertheless, considerable ongoing debate regarding the true yield of the Trinity test exists in the literature, presumably enabled by (a) ready access to trinitite and high-resolution gamma ray spectrometers and (b) the fairly large uncertainty inferred by the differences in official DOE yield quotations over time. The latter is significant, as one would assume that the Laboratory that fielded the test would have the most accurate and stable interpretation of performance. Reality is, as just described, complicated by the chemical variability in the original dataset. Further, when one considers the fairly short half-lives of the fission fragments that are chemically suitable and diagnostically useful (typically days to a few months), it becomes obvious that there was no way to go back and re-measure them for Trinity, even just a few years after the test. Various LANL efforts throughout the years have refined the assessments – hence the updated yield statements – but all of these are ultimately limited by the original dataset. Many of the open literature studies emphasize longer half-life fission products like $^{137}$Cs ($t_{1/2}$ = 30 y), but these species do not behave chemically like plutonium. This is demonstrated by widely varying activity levels sample-to-sample and correspondingly large ranges in yield estimates.[11] All of which might suggest that Trinity's yield cannot be more accurately known than it was in 1945.

### III. Extinct Radionuclides

Recently, Los Alamos National Laboratory's (LANL) Nuclear and Radiochemistry Group (C-NR, the organizational descendant of the original Radiochemistry Group) initiated studies aimed specifically at "resurrecting" the very fission fragments originally used for determining plutonium efficiency in trinitite. This seemingly impossible idea was made possible by the fact that the original fission fragments had not "disappeared." Rather, they simply decayed into stable elements. These stable decay daughters are referred to as extinct radionuclides because they are no longer radioactive, and are therefore no longer detectable radiometrically. However, once fixed in a piece of solid trinitite, the short-lived fission fragments quantitatively decay up their beta chains into stable isotopes of another element. This subtly perturbs the natural abundances of the chain terminus isotopes in the vitreous solid. A cartoon of this process is shown in Figure 1. The magnitude of the fission-induced perturbation is minute, typically on the order of a few hundredths of a percent. Until the last decade or so, the precision required to measure this perturbation was not commonly available. Today, high precision mass spectrometric instruments are widespread and LANL successfully employed them to quantify fission-induced $^{95,97}$Mo$_{Zr}$ in trinitite. The subscript "Zr" refers to the elemental state of the original short-lived radionuclide, with molybdenum being the chain terminus element for the 95- and 97-mass fission fragments. This work was reported by Hanson et al. in the July 2016 edition of the Proceedings of the National Academy of Science.[5]

| | $^{92}$Mo 14.84 | $^{93}$Mo 3.5E3a | $^{94}$Mo 9.25 | $^{95}$Mo 15.92 | $^{96}$Mo 16.68 | $^{97}$Mo 9.55 | $^{98}$Mo 24.13 | $^{99}$Mo 2.75d | $^{100}$Mo 9.63 | | | |
|---|---|---|---|---|---|---|---|---|---|---|---|---|
| | | $^{92}$Nb 3.5E7a | $^{93}$Nb 100 | $^{94}$Nb 2.0E4a | $^{95}$Nb 34.99d | $^{96}$Nb 23.4h | $^{97}$Nb 1.23h | $^{98}$Nb 2.9s | $^{99}$Nb 15.0s | $^{100}$Nb 1.5s | | |
| | | | $^{92}$Zr 17.15 | $^{93}$Zr 1.5E6a | $^{94}$Zr 17.38 | $^{95}$Zr 64.02d | $^{96}$Zr 2.8 | $^{97}$Zr 16.8h | $^{98}$Zr 30.7s | $^{99}$Zr 2.2s | $^{100}$Zr 7.1s | |
| | | | | $^{92}$Y 3.54h | $^{93}$Y 10.2h | $^{94}$Y 18.7m | $^{95}$Y 10.3m | $^{96}$Y 5.3s | $^{97}$Y 3.76s | $^{98}$Y 0.59s | $^{99}$Y 1.47s | $^{100}$Y 0.73s |
| | | | | | $^{92}$Sr 2.71h | $^{93}$Sr 7.41m | $^{94}$Sr 1.25m | $^{95}$Sr 25.1s | $^{96}$Sr 1.07s | $^{97}$Sr 0.43s | $^{98}$Sr 0.65s | $^{99}$Sr 269ms | $^{100}$Sr 201ms |

Figure 1. Partial scheme for the decay of fission products into stable molybdenum and zirconium species. Stable species are denoted with gray boxes, and the natural isotopic abundance is shown for each stable species. Radioactive species and their half-lives are shown in white boxes. Blue and red arrows illustrate β-decay chains of interest, and the bolded blue and red boxes indicate stable isotopes that are perturbed by decay of short lived fission products.

The Proceedings article was intended to demonstrate the similarity of chemistries between plutonium and zirconium upon condensation, as compared to more readily measured $^{137}$Cs. Evidence of plutonium/zirconium co-condensation was made clear by the very consistent ratios of F and Pu$^{res}$, also measured in the same samples. As a validation of the hypothesis, the plutonium efficiency was calculated using Equation 3 and the corresponding yield (Equation 4) was compared to the official DOE Trinity yield. Good agreement was observed, and is updated below. Although the study was focused on proving a chemical phenomenology hypothesis, it suggested that extinct radionuclides might provide the opportunity for the first complete study of Trinity's yield since 1945.

Unlike the Proceedings study, the new effort was aimed at fully understanding the Trinity device's performance. To this end, thirteen samples of trinitite were analyzed for a suite of isotopes, some measured using traditional radiometric techniques, some via the newly developed extinct radionuclide methods. Constructing an accurate picture of Trinity's performance employed what is referred to as the Extinct Radionuclide System (ERS), which folded the measurement data with modern radiochemical and physics simulation tools. The system also included a Bayesian statistical treatment that was developed in parallel with the extinct radionuclide measurements. This ensured a rigorous uncertainty statement for the ERS assessment of Trinity's yield, $24.8 \pm 2$ kt.[12]

## IV. Conclusion

Now, for the first time since the original measurements were made in 1945, we can report a new yield for Trinity derived solely from original trinitite samples, using modern experimental and analytical methods. The ERS study shows that Trinity's yield is $24.8 \pm 2$ kt TNT equivalent, somewhat higher than current and historic estimates. This new assessment's uncertainty overlaps with the previous DOE value of $21 \pm 2$ kt, where we assigned a conservative 10% relative uncertainty on its value of 21 kt. The source of the accepted DOE yield is not cited in the official DOE publications. However, recently discovered internal communications suggest that the updated yield was the work of Charles I. Browne, leader of the Radiochemistry Group in 1963.[13] The memos are unambiguous regarding the fact that yield assessment was still limited by the original dataset. Browne informed his reassessment of Trinity with data from the similar Crossroads tests fired at the Bikini atoll in 1946. While a reasonable approach, the differences in location, height of burst, and myriad other environmental and engineering factors relative to Trinity are all troubling confounds. This suggests Browne's yield is likely less well known than our 10% uncertainty estimate would indicate. In contrast, the present assessment is based on internally consistent data from trinitite samples, all analyzed in the same manner. The data has been evaluated using the most sophisticated modeling tools available, and the resultant models and data subjected to statistical scrutiny. We believe the ERS yield of Trinity is the most accurate and comprehensively characterized since the original Los Alamos Radiochemistry Group's first effort. We suggest that $24.8 \pm 2$ kt now be adopted as the yield of the Trinity event.

## V. Acknowledgments

The authors gratefully acknowledge the support of the Los Alamos National Laboratory's Laboratory Directed Research and Development (LDRD) investment 20160011DR. We also enjoyed many helpful discussions and edits from M. Chadwick, C. Carmer, K. Spencer, D. Hollis, L. Drake, and D. Mercer. A special thanks is due to B. Archer for his discovery of the memoranda that led to the official DOE yield of Trinity.

*This work was supported by the US Department of Energy through the Los Alamos National Laboratory. Los Alamos National Laboratory is operated by Triad National Security, LLC, for the National Nuclear Security Administration of the US Department of Energy under Contract No. 89233218CNA000001.*